# A Model of UV-Blue Absorbance in Bulk Liquid of Venusian Cloud Aerosols Is Consistent with Efficient Organic Absorbers at High Concentrations

Jan Spacek[1], Paul B. Rimmer[2], Janusz J. Petkowski[3,4], and Yeon J. Lee*[5]

[1]Foundation for Applied Molecular Evolution, 13709 Progress Boulevard, Alachua, FL 32615.
[2]Cavendish Astrophysics, University of Cambridge, Cambridge, CB3 0HE, UK.
[3]Faculty of Environmental Engineering, Wroclaw University of Science and Technology, 50-370 Wroclaw, Poland.
[4]JJ Scientific, Mazowieckie, 02-792 Warsaw, Poland
[5]Planetary Atmospheres Group, Institute for Basic Science (IBS), Daejeon, South Korea.
Corresponding author: Yeon Joo Lee email: yeonjoolee@ibs.re.kr

# Abstract

At visible wavelengths, Venus appears serene and pale yellow. But since the 1920s, observers have noted high-contrast features in the ultraviolet. These features track the ~4-day superrotation of the upper cloud deck and vary widely over time and space. The identity of the UV absorber(s)—active between at least 280 and 500 nm—remains unknown, as no proposed candidate fully matches all observational data. From remote observations of Venus, and accounting for light scattering by cloud droplets, we modeled the 365–455 nm decadic absorption coefficient, $a_\lambda$, of the bulk liquid that form Venus's clouds. Assuming a uniform distribution in mode 1 and 2 particles across a 6 km layer below the cloud top at 65 km, we constrained the decadic absorption coefficient within the modeled range to a peak at $a_{375}$ = 1278 cm$^{-1}$, equivalent to a decadic absorbance of $A_{375}$ = 1278 for a 1 cm path length. This extremely high absorption coefficient implies the presence of a highly efficient absorber, for example, conjugated organics, at relatively high concentration—for example, ~12 g/L for porphyrin-type pigments with a representative peak molar absorption coefficient of ~10$^5$ M$^{-1}$cm$^{-1}$. Inorganic absorbers, typically below 10$^4$ M$^{-1}$cm$^{-1}$, would either need to constitute a large portion of the aerosols or still not be sufficiently light-absorbing, even in pure form. We emphasize that all candidate absorbers must be evaluated against Venus's reflectance curve using (i) known molar absorption coefficients, (ii) realistic atmospheric distributions, and (iii) appropriate particle size distributions. The planned Rocket Lab mission will test the hypothesis of organics in Venus's clouds.

Key words: Venusian clouds, UV absorber, Organic pigments, Aerosol modeling, Radiative transfer.

# 1. Introduction

To the naked eye and to amateur telescope hobbyists, Venus appears to be pale yellow. However, ground-based astronomy a century ago began to create a more sophisticated set of observations. These early observations in the ultraviolet (UV) revealed dark streaks and patches on the venusian disk (Ross, 1928, Wright, 1927).

The reflected brightness of Venus is shaped by (1) incoming solar radiance, (2) scattering by atmospheric molecules and aerosol particles, and (3) absorption by gases and aerosol particles, located predominantly in the global cloud layer. Investigation of the reflected spectrum provides essential clues to understand optical properties of atmospheric gases and aerosols remotely, allowing us to suggest the most likely chemical composition. Utilization of such a reflected spectrum of Venus has been intensively conducted with space-based measurements from ultraviolet to near-infrared wavelengths and radiative transfer numerical calculations that include known atmospheric conditions.

The cloud-top albedo (and associated radiant brightness) has been extensively studied in the 350–460 nm range (Lee et al., 2022). These studies, and studies prior to them, reveal that the unexplained absorption likely extends down to at least 280 nm (Pollack et al., 1980) and potentially beyond 500 nm (Pérez-Hoyos et al., 2018).

After a century of observations, the substance or group of substances responsible for this phenomenon is still unknown; the venusian absorber is, in fact, referred to in the literature as the "unknown absorber" (Barker et al., 1975, Mills et al., 2007, Pollack et al., 1980, Titov et al., 2018, Zasova et al., 1981). Over the years, more than 20 different chemicals have been proposed as the source of the absorber. These candidates include gas-phase molecules such as OSSO (Frandsen et al., 2016), $S_2O$ (Na and Esposito, 1997, Pérez-Hoyos et al., 2018), molecules in the cloud aerosol such as $S_x$ (Toon et al., 1982), $Cl_2$ (Pollack et al., 1980), ferric iron (Krasnopolsky, 2018, Petrova, 2018, Zasova et al., 1981), and iron-sulfur minerals (Jiang et al., 2024), to name a few. Each of these candidates has issues (Egan et al., 2025a, Krasnopolsky, 2016, Pérez-Hoyos et al., 2018). For a summary of the proposed candidates, see the works of Limaye et al. (2021), Mills et al. (2007), Pérez-Hoyos et al. (2018), Petkowski et al. (2024).

*In situ* photometric measurements show that the unknown absorption occurs predominantly just below the cloud tops. They also show an absence of sharp spectral features at 0.3–0.4 μm, which strongly indicates that the unknown absorber is associated with the cloud particles rather than gaseous species (Ekonomov et al., 1984, Titov et al., 2012, Tomasko et al., 1979, Tomasko et al., 1980), reviewed by (Titov et al., 2018). However, we note that the exact contributions of aerosols and gases to the total solar light absorption within the clouds is a matter of debate and still being explored. *In situ* measurements provide strong evidence that the most significant solar energy absorption occurs in the upper cloud layer (Ekonomov et al., 1984, Tomasko et al., 1980). Similarly, the phase angle sensitivity, evident in albedos derived from remote sensing observations of the venusian cloud top radiance, for example, glory phenomena, cannot be explained solely by gas absorption or Rayleigh gas scattering but can be explained by Mie scattering sensitivity to aerosol properties (Lee et al., 2017, Markiewicz et al., 2014, Markiewicz et al., 2018, Muñoz et al., 2014).

Finding a good candidate that would explain the unknown absorber is difficult due to poorly defined details of the absorber's spectrum. The spectrum profile is well-defined only for the 350–450 nm range (Lee et al., 2022), which is a fraction of the spectral range covered by the absorber. Some observations suggest that the absorber is active from at least 200 nm (Marcq et al., 2020) to 600 nm (Pérez-Hoyos et al., 2018). More recent Akatsuki observations, however, indicate considerable absorption primarily over the 350–450 nm range, with absorption at longer wavelengths remaining weakly constrained (Lee et al., 2022). MESSENGER's MASCS observations suggest that it may have a maximum at 340 nm, with a full width at half maximum (FWHM) of 140 nm (Pérez-Hoyos et al., 2018). The MESSENGER mission could not look at wavelengths below 300 nm. Spectral data at shorter wavelengths were acquired by the SPICAV spectrometer aboard Venus Express and covered the spectral range of 170–320 nm.

The observed cloud-top albedo undergoes significant spatial and temporal changes. This variability has been attributed to spatial and temporal variations in the unknown absorber's abundance, using spectral ranges that do not contain impacts of known atmospheric gases (Crisp, 1986, Lee et al., 2019, Pérez-Hoyos et al., 2018, Titov et al., 2012). Although the unknown absorber is usually presumed to be a single chemical species, it may be linked to multiple absorbers. Cottini et al. (2012) presented no clear correlation between the cloud-top altitudes and the unknown absorber's abundance, implying that we lack observational clues to consider vertical distribution changes of the unknown absorber. Even though the Hubble Space Telescope provides estimates of UV albedo across wavelengths from 200 to 550 nm on the same date and same spatial location (Jessup et al., 2020), the spatial and temporal variations are not due to topographic impacts (Lee et al., 2020b).

Temporal changes in the absorption required to match the observed cloud-top brightness at 365 nm and near 250 nm correlate, over both short- and long-term periods, with $SO_2$ gas abundance near the cloud tops (Del Genio and Rossow, 1982, Esposito et al., 1988, Imai et al., 2019, Lee et al., 2015a, Lee et al., 2020a, Lee et al., 2019, Marcq et al., 2013, Marcq et al., 2020). These correlations may link volcanic outgassing of $SO_2$ to photochemical processes that, in turn, can affect cloud formation (Mills et al., 2007). The short-term variations suggest global-scale atmospheric waves with a 4-5-day periodicity and varying amplitudes (Del Genio and Rossow, 1982, Imai et al., 2019, Lee et al., 2020a). Long-term variations in the absorber's intensity, occurring on decadal timescales, can also significantly affect Venus's global atmospheric processes, including possible upwellings that transport the putative absorber from lower atmospheric layers (Bertaux et al., 1996).

The unknown absorber is remarkably efficient. It captures more than 50% of the solar energy that reaches the cloud-top atmosphere of Venus (Crisp, 1986, Lee et al., 2015b), with consequent effects on atmospheric structure and dynamics, in particular changes in global-scale atmospheric circulation and zonal wind speeds (Lee et al., 2019).

Largely absent from this discussion, with a few exceptions (Hartley et al., 1989), is the possibility that the unknown absorber is organic (typically a molecule with C-H or C-C bond, irrespective of biological origin). For example, Dayhoff and Sagan write that "[t]here can be no free carbon, hydrocarbons, formaldehyde, or any other organic molecule present in more than trace amounts [in the clouds of Venus]" (Dayhoff et al., 1967). Larger organics have been thought to be incompatible with the high acidity of the venusian clouds (Marov, 1972). Such views have influenced modeling of the venusian atmosphere to "not account for the existence of reduced carbon compounds" (Code, 1972, Titov et al., 2018).

In a theoretical paper (Spacek, 2021), Spacek reconsiders the possibility that organics may be observationally significant for Venus. His model for the chemical change of organics in venusian $H_2SO_4$ relies on studies from the petrochemical industry, which has long been interested in the reactivity of organic molecules in $H_2SO_4$ (Albright et al., 1971). The model suggests that, contrary to popular belief, organics in highly concentrated sulfuric acid (for w/w in water > 75%) are not oxidatively destroyed to yield small volatile molecules lacking chromophores in the near UV and visible spectral regions. Instead, organics in concentrated sulfuric acid are transformed to larger and more complex multi-carbon species. These mixtures are called acid-soluble oils (ASO), conjunct polymers (Xin et al., 2022), or prosaically "red oil," based on their visual properties (Albright et al., 1971).

Motivated by this model, Spacek et al. (2024) examined chemical processes transforming simple organic molecules in highly concentrated sulfuric acid into complex molecules that could in principle be present in the clouds of Venus and that could potentially give rise to the unknown absorber. These experimental studies showed the first steps in the chemical processes that might lead to the formation of "red oil" in the clouds of Venus. In detail, gases believed to be present in Venus's atmosphere, formaldehyde (HCHO) and carbon monoxide (CO), react in highly concentrated sulfuric acid to yield glycolic acid ($HOCH_2COOH$) (Spacek et al., 2024). While glycolic acid is remarkably stable in sulfuric acid at concentrations represented by the venusian aerosol droplets, it slowly (weeks at room temperature, hours at 100°C) transforms to more complex fluorescent and UV-absorbing organic species. The absorption profile of the organic mixture arising from HCHO and glycolic acid treated in sulfuric acid (Spacek et al., 2024) is similar to the spectrum of graphene oxide (Johra et al., 2014) or crude oil (Banda-Cruz et al., 2017).

Such chemical reactions were proposed to happen in venusian clouds as well, where a series of chemical transformations, analogous to the formation of microporous humins (Björnerbäck et al., 2018), could eventually form carbon-rich particles at the bottom of the clouds. Such carbon-rich particles would be resistant to evaporation or cracking (splitting of aliphatic hydrocarbon chains). After falling from the cloud to higher temperatures nearer to the venusian surface, the carbon-rich particles would react in a Boudouard reaction with $CO_2$ to give CO at lower altitudes (<40 km) (Mianowski et al., 2015), completing the proposed carbon cycle (Spacek, 2021). The CO would be fully oxidized near the surface, likely in proportions consistent with the surface oxidation state (Constantinou et al., 2024, Fegley Jr et al., 1997).

Highly concentrated sulfuric acid solutions holding organics usually range from colorless to pale yellow to black, depending on the concentration of organics and age of the solution. Older mixtures turn darker, as larger light-absorbing conjugated systems are formed (Moncada et al., 2024). However, note that when observed by scattered sunlight, the apparent color of a cloud material is determined by a combination of particle size distribution and the absorbance properties of the bulk material. Hence the observed color of the cloud is determined by a combination of backscattering of incident light from the particles and absorption within the particles, while the color of the bulk liquid is generally dominated by the absorption of light within the material. The difference increases with the surface-area-to-volume ratio of the absorbing material; the smaller the particle, the higher the scattering. This phenomenon can be illustrated with an example. Imagine a waterfall. While the bulk water in the river may appear dark with impurities, as it breaks down to smaller particles during the fall it gets gradually brighter, with a cloud of fine mist around the bottom appearing white. Thus, observing a cloud of aerosols reveals little about the optical properties of the bulk liquid it contains without knowledge of the particle size distribution and a model that accounts for light scattering.

In laboratory measurements, the absorption of bulk solutions correlates with the concentrations of dissolved chromophores (Beer's law) (Swinehart, 1962). To correlate our experimental studies of sulfuric acid solutions with the planetary observables, we calculated, for the first time, the linear decadic absorption coefficient of the bulk liquid forming Venus's cloud aerosol droplets over the 365–455 nm range. We assumed that the UV absorber is in the cloud particles, and not in the gas phase. Under this assumption, we constrained the absorbance properties of Venus's cloud particles required to explain the UV/vis absorber, as a function of particle size distributions. We found that the bulk liquid that forms venusian clouds must be highly absorbing in the 365-455 nm region, with a steep drop-off toward 455 nm. We show that some organic compounds are viable candidates for the UV absorber, solely in terms of their optical properties, with predictive power for future experiments and space missions.

In Section 2.1, we show how, from the astronomical observations, we modeled the absorbance spectrum of the bulk liquid that forms Venus's clouds. In Section 2.2, we constrain the structure and concentration of the absorber under the assumption that the absorber is an organic molecule with a defined structure or a mixture of specific organic molecules. We conclude the paper with the discussion and summary (Section 3).

# 2. Results

## 2.1. Modeling the absorbance of the liquid in Venus’s clouds

Identifying and characterizing a candidate absorber in the venusian clouds require matching astronomical observations of the unknown absorber with laboratory studies of the absolute spectroscopic profile of a bulk liquid. Such comparisons between astronomical observations and laboratory measurements are not trivial and cannot be correlated directly, as one needs to account for effects due to scattering and absorption present in both observations to different degrees. In laboratory measurements, the light absorption within solutions dominates the observed optical properties to the point that the effects due to light scattering are typically neglected (Swinehart, 1962). By contrast, observations of light reflected from the cloud tops are dominated by scattering, and the light absorption within the bulk material of the sulfuric acid aerosols has negligible effect at the visible and shorter wavelength range. Because scattering dominates the optical properties of the reflected light in the aerosols, we cannot determine the absorption intensity profile (color) of the bulk liquid from the apparent (reflected) color of the cloud without a radiative transfer model that utilizes the known particle size and spatial distributions. The contribution of scattering relative to absorption is more significant for smaller particles.

Some authors attempted to adjust cloud-aerosol optical properties to explain the observed spectra of Venus. For example, while the sulfuric acid cloud aerosols are effectively non-absorbing and dominated by scattering at wavelengths less than 600 nm, some studies increased the imaginary index of an aerosol or the ratio of absorption to extinction (the sum of scattering and absorption) (Lee et al., 2022, Marcq et al., 2020, Petrova, 2018, Pollack et al., 1980). Others directly correlated the observed venusian albedo with the absorbance of bulk solutions (Egan et al., 2025b, Jiang et al., 2024, Limaye et al., 2018). However, none of the previous studies distinguished scattering and absorption to calculate the absolute absorption spectrum of a sulfuric acid aerosol mixture that contains absorbing solutes.

Our model uses data from remote observational studies of reflectivity spectra acquired over the entire venusian day-side in 2020 (Lee et al., 2022). We analyzed the observed spectra using radiative transfer modeling to determine the linear decadic absorption coefficient of the bulk liquid present in aerosols in Venus’s upper cloud. The model considers multiple scattering by the cloud aerosols and atmospheric molecules, as well as the latest data on the absorption of the atmospheric gases and their vertical distribution, specifically, $SO_2$, OCS, $O_3$, $CO_2$, SO, $H_2O$, $H_2S$, and HCl (Lee et al., 2022). For example, we take the abundance of $SO_2$ to be 450 ppbv at 61 km altitude, an average of the ground-based measurements from 2016 to 2018. The $SO_2$ gas abundance decreases upward by the 3 km scale height of $SO_2$ gas (Marcq et al., 2020), which results in 22.4 ppbv near the cloud-top level (70 km). The vertical distribution of the $O_3$ gas near the cloud-top level follows the work of Marcq et al. (2019). Note that the absorption of the gaseous components of the atmosphere plays an important role at wavelengths shorter than ~320 nm but is insufficient for longer wavelengths. Above ~320 nm the assumption of the unknown absorber present within the aerosols is necessary to explain the observed venusian cloud-top albedo. Our model uses a cloud-top structure that decreases exponentially upwards to take into account the changes in the density of cloud particles with altitude.

We take the observationally constrained size distribution of the venusian cloud aerosols (see Table 1) and observed reflectance curve from 365 to 455 nm (Lee et al., 2022). We determine what range of wavelength-dependent absorption is required beyond all of the known components, which implies the contributions of the unknown absorber.

The size of cloud aerosols near the cloud-top level (~65 km) and above can be estimated using remote sensing techniques, such as glory optical phenomenon analysis (Lee et al., 2017, Petrova et al., 2015) and spectral dependence of atmospheric extinction with solar occultation measurements (Luginin et al., 2016). Such observational data analysis uses the reflected or scattered solar radiance by the clouds and the upper haze, complementing limited information at that altitude with *in situ* measurements (Knollenberg and Hunten, 1980). The results of global mean aerosol sizes in Lee et al. (2017) are close to those in Petrova et al. (2015), measured between 35°S and 62°S latitudes. The aerosol size may decrease with increasing altitude up to 85 km (Luginin et al., 2016).

*Table 1. Summary of the effective radius ($R_{eff}$) and effective variance ($v_{eff}$) of the mode 1 and mode 2 particles.*

| References | $R_{eff1}$ [μm] | $v_{eff1}$ | $R_{eff2}$ [μm] | $v_{eff2}$ |
|---|---|---|---|---|
| Pollack *et al.* (1980) | 0.43 (log-normal, $\bar{r}$=0.15) | 0.52 (log-normal, $\sigma$=1.91) | 1.15 (log-normal $\bar{r}$=1.05) | 0.04 (log-normal, $\sigma$=1.21) |
| Kawabata *et al.* (1980) | 0.23 | 0.18 | | |
| Petrova *et al.* (2015) | 0.23 | 0.18 | 1.05–1.2 | 0.07 |
| Luginin *et al.* (2016) | 0.12 ± 0.03 | 0.15 | 0.84 ± 0.16 | 0.1 |
| Lee *et al.* (2017) | 0.43* | 0.52* | 1.26 | 0.076 |
| **Used in our model** | **0.43** | **0.52** | **1.26** | **0.076** |

*Referring Pollack *et al.* (1980)

To understand how aerosol size distribution affects mean extinction cross-section per particle, we compared the possible range of sizes in Fig. 1 for mode 1 and mode 2. The extinction cross-section varies depending on the size distributions of particles, which were retrieved using recent remote-sensing observations (Lee et al., 2022, Pérez-Hoyos et al., 2018). This study assumed the refractive index for 75% sulfuric acid aerosols (Hummel et al., 1988) in the Lorenz-Mie code calculations (Mishchenko et al., 2002) with log-normal size distributions. We note that there are considerable impacts of size assumptions. In this study, we used the global mean value: mode 1 with $r_{eff}$ = 0.43 μm and $v_{eff}$ = 0.52 (Pollack et al., 1980), and mode 2 with $r_{eff}$ = 1.26 μm and $v_{eff}$ = 0.076 (Lee et al., 2017), as shown in the red dots in Fig. 1.

In the cloud model, we assume the ratio of extinction coefficients between the two modes to be one at 365 nm, following the *in situ* measurement results by the Pioneer Venus Sounder Probe (Knollenberg and Hunten, 1980, Pollack et al., 1980); also, the total extinction coefficient of clouds changes vertically with the 4-km cloud scale height (Ignatiev et al., 2009, Lee et al., 2012, Satoh et al., 2015). We assume that the cloud-top altitude is 65 km, the unknown absorber is well-mixed between the two-mode aerosols and its vertical distribution is limited to a 6-km-thick layer right below the cloud tops, from 59 to 65 km, which was a crucial vertical location to explain the observed phase curve of Venus at 365 nm (Lee et al., 2021) and also to be consistent with the sharp increase of absorption or shallow layer of absorption found in previous studies (Esposito, 1980, Pollack et al., 1980). The upper boundary of the absorbing layer can be higher, or the lower boundary of the layer can be deeper, while the crucial condition to explain the observed phase curve is the absorption by a layer right below the cloud tops with a thickness of one cloud scale height. Importantly, we are not suggesting that the UV absorber is relegated to the upper clouds, only that the reflectance data we used to constrain the absorber is only sensitive to the upper cloud region. The unknown absorber is demonstrated by reducing the single scattering albedo (SSA) of the cloud aerosols. SSA of a particle is the ratio of scattering efficiency to total extinction efficiency, SSA = scattering / (scattering + absorption). SSA = 1.0 implies non-absorbing particles, and SSA < 1.0 implies absorbing particles. Reducing SSA mimics increasing the imaginary index of the 75% w/w $H_2SO_4$-$H_2O$ liquid aerosols. Note that the imaginary index of the sulfuric acid aerosols is $1.07 \times 10^{-8}$ at the wavelengths shorter than 550 nm and increases to $1.06 \times 10^{-5}$ at 1.3 µm. Despite this increase, the imaginary index remains small, and the calculated SSA remains close to 1.0 at wavelengths shorter than 1.3 µm (Hummel et al., 1988).

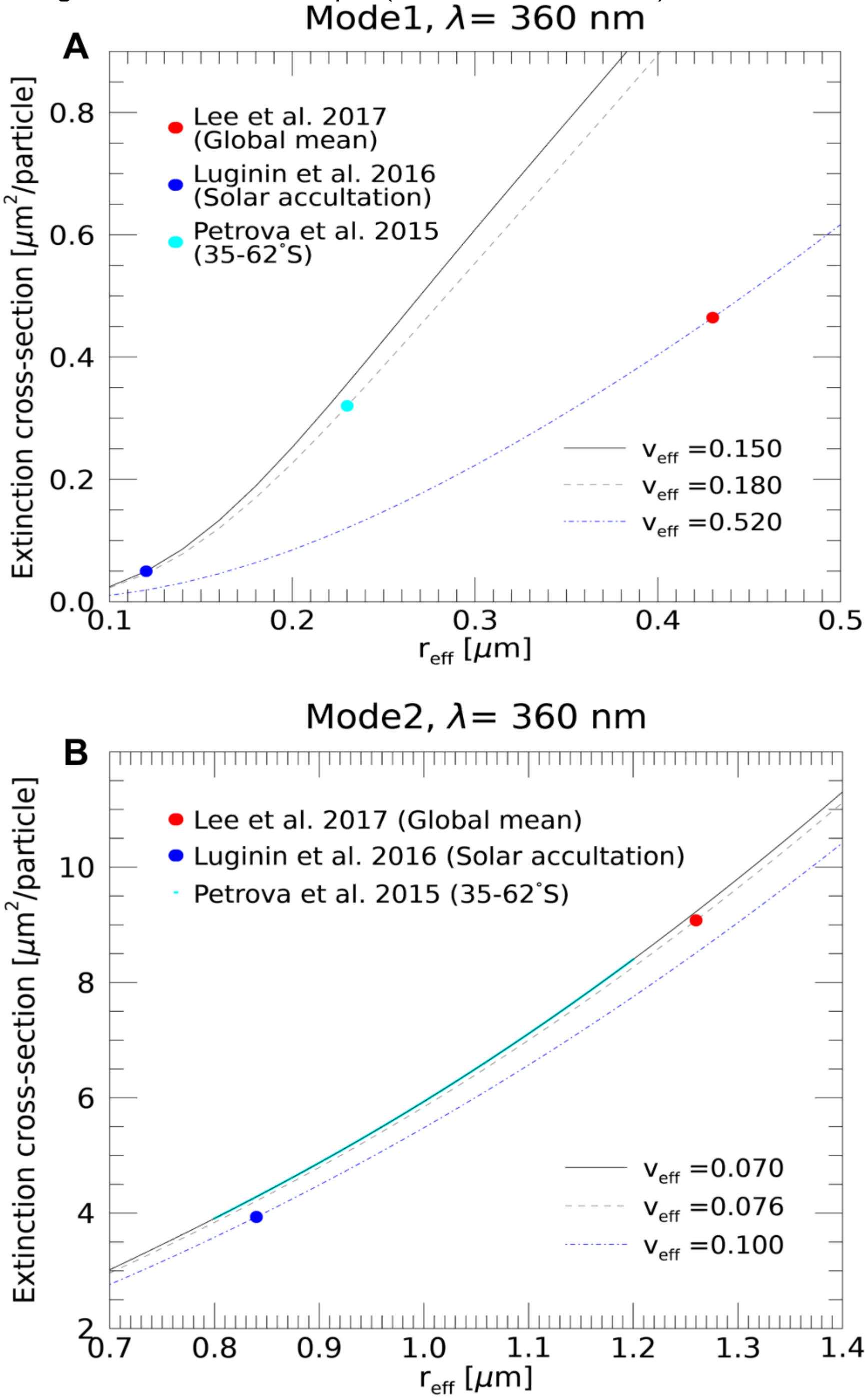

***Fig. 1.*** *Mean extinction cross-section per particle at 360 nm. The red dots represent data used in this study (Table 1) (Hummel et al., 1988, Lee et al., 2022, Mishchenko et al., 2002, Pérez-Hoyos et al., 2018). Graph A: For small particles (mode 1), the effective variances ($v_{eff}$) are fixed at 0.15,*

*0.18, and 0.52. Effective radii ($r_{eff}$) are changed in the range of 0.1–0.5 μm. Graph B: For large particles (mode 2), the effective variances are fixed as 0.070, 0.076, and 0.100, while effective radii are set in the range 0.7–1.4 μm. The results of remote-sensing data analysis are indicated by three colored points or a cyan line (bottom panel).*

The required SSA of the cloud aerosols is 0.89 at 365 nm for both mode 1 and mode 2 particles to reproduce the 2020 Venus day-side campaign data. By definition, an SSA this close to unity indicates that scattering dominates over absorption at this wavelength. An SSA of 0.89 corresponds to the imaginary index of 0.01 for the mode 1 particle and 0.003 for the mode 2 particle. The inferred imaginary index for the mode 1 particle is comparable to previous studies near 360 nm (~0.02) (Pérez-Hoyos et al., 2018) and within the range reported at 250 nm (0.001–0.5) (Marcq et al., 2020). Using the required SSA of the cloud aerosols and the effective radii of the cloud particles, we calculate the linear Napierian absorption coefficient:

$$\alpha_\lambda = \frac{\sigma_{abs}(\lambda)}{V_{ptcl}}$$

where $\alpha_\lambda$ is the wavelength-dependent linear Napierian absorption coefficient, $\sigma_{abs}$ is the absorption cross-section per particle, and $V_{ptcl}$ is the volume of a cloud aerosol particle.

For an optical path length $l$, the corresponding Napierian absorbance is:

$$B_\lambda = \alpha_\lambda l$$

The linear decadic absorption coefficient, $a_\lambda$, is obtained from the linear Napierian absorption coefficient by:

$$a_\lambda = \frac{\alpha_\lambda}{\ln(10)} = \frac{\sigma_{abs}(\lambda)}{\ln(10) V_{ptcl}}$$

Accordingly, the decadic absorbance over the same optical path length is:

$$A_\lambda = \frac{B_\lambda}{\ln(10)} = \frac{\alpha_\lambda l}{\ln(10)} = a_\lambda l \ = \varepsilon_\lambda c l$$

The decadic absorbance, $A_\lambda$, is the quantity measured in the laboratory by UV–visible absorbance spectroscopy. It depends on the wavelength-dependent molar absorption coefficient of the absorbing molecule or mixture, $\varepsilon_\lambda$ ($M^{-1}$ $cm^{-1}$), its molar concentration, $c$ (M), and the optical path length, $l$ (cm). Below, we describe the model used to constrain the complex refractive index and, from this, the concentrations of absorbers with known molar absorption coefficients.

Figure 2 shows the calculated decadic absorption coefficient, $a_\lambda$, as a function of wavelength.

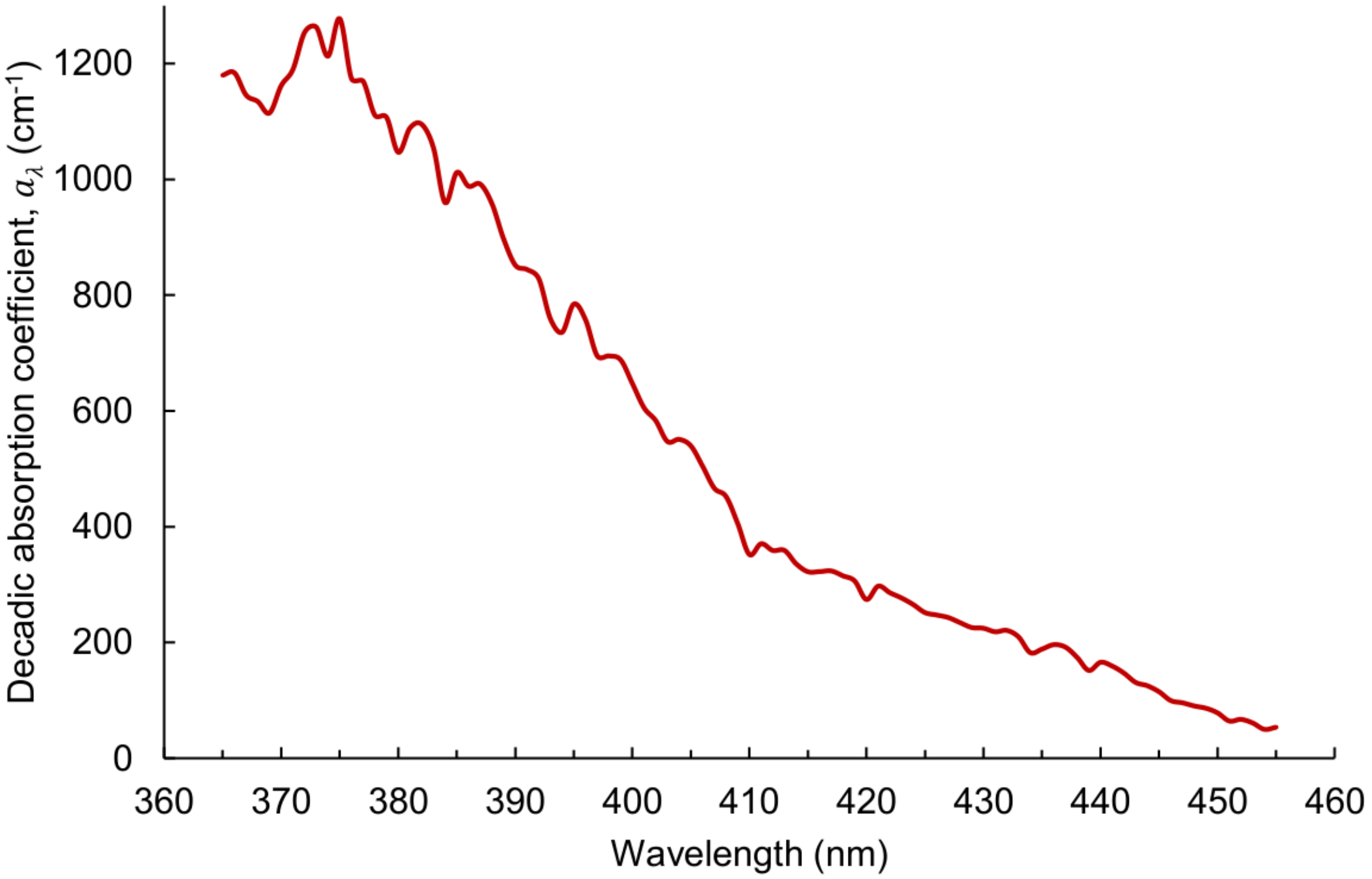

***Fig. 2**. The average decadic absorption coefficient, $a_\lambda$, of the bulk liquid comprising Venus's cloud aerosols. The curve (red) is back-calculated from observations of Venus's atmosphere using a multiple-scattering radiative-transfer model and assumptions about the particle size distributions and number density. The coefficient reaches 1278 cm$^{-1}$ at 375 nm, which implies the presence of a high concentration of highly absorbing species within Venus's cloud aerosols (see Table 2). For a 1 cm path length, $a_\lambda$ is numerically equal to $A_\lambda$, although $A_\lambda$ is dimensionless.*

Although the UV/vis absorber was suggested to be active from at least 283 nm (Lee et al., 2021) to "well past 0.5 μm" (Pérez-Hoyos et al., 2018), our model is limited to the 365–455 nm part of the spectrum, because it is not clear whether above-zero absorption at wavelengths longer than 450 nm is statistically significant.

## 2.2. Comparing the model to organic absorbers

The UV-vis absorbance spectrum, which in the visible-wavelength range influences the perceived color, can be used to constrain the structure of an organic absorber. The linear decadic absorption coefficient $a_\lambda$, which influences how strongly a bulk liquid absorbs light at a given wavelength, provides insights into the compound's concentration.

### 2.2.1. The absorption spectrum suggests the structure of the absorber(s)

Photons within an appropriate energy range may be absorbed by organic molecules, exciting one or several electrons to higher orbitals or causing an electron transition between chemical species. Both the photon energy value and energy range required for the electron excitation inform us about the structure of the absorber. For the near UV to visible wavelength range, the most common absorption is due to $\pi \rightarrow \pi^*$ electronic transitions.

In general, the larger the conjugated system, the smaller the energy gap between the highest occupied molecular orbital (HOMO) and the lowest unoccupied molecular orbital (LUMO), resulting in absorption of light with longer wavelengths. This observation was formalized with Woodward's empirical rules, where each additional functional group adds to the base absorption wavelength by a specific value (Kalsi, 2007). The absorption wavelength can also be modeled using time-dependent density functional theory, usually resulting in more precise predictions of the spectrum profile based on the known structure (Chen et al., 2020, Zuehlsdorff and Isborn, 2019). However, both of these methods become less precise as the size of the system increases, especially if a transition/exchange method other than π→π* electronic transitions (such as charge-transfer) is involved. Note that the models cannot be used to provide the exact quantum spectral structures, as various solutes might give closely matching absorption spectra.

The part of the spectrum we modeled in this work does not fully cover the entire absorption spectrum of the unknown absorbers as reported previously (Pérez-Hoyos et al., 2018); however, it covers a sufficiently wide absorption wavelength range (365–455 nm). The key aspect is the absolute reflectivity retrieval using observations to provide the required absorption at this wavelength range.

If the absorber is organic, then the 340 nm absorption retrieved by Pérez-Hoyos et al., (2018) most likely arises from π→π* transitions in a given conjugated organic molecule. The candidate is likely a moderately sized conjugated system (larger than modified naphthalene but smaller than extensive graphene-like systems). The broad signal (FWHM =140 nm, with absorption extending from ~280–500 nm (Pérez-Hoyos et al., 2018)) may suggest a charge-transfer interaction or a flexible molecule. The candidate is likely not a single well-defined rigid molecule with sharp distinct energy levels occurring solely due to π→π* transitions; however, it can be a mixture of such molecules with overlapping absorption maxima.

This absorption profile narrowed the candidate pool of organic absorbers; however, many organic molecules, or molecule mixtures, might still fit the profile. As an example, a mixture of quinine derivatives might fit the spectrum. Quinine itself in 1 M sulfuric acid has $\lambda_{max}$ = 349 nm; FWHM = 72 nm (Irvin and Irvin, 1948), with a second, more prominent peak around 250 nm. If we were to consider a single organic molecule as a candidate for the unknown venusian absorber, a large porphyrin derivative akin to bacteriochlorophyll from *Blastochloris viridis* (Fig. 3) would fit well the spectral absorption behavior within the modeled section of the unknown absorber spectrum (Pucelik et al., 2020).

## 2.2.2. Constraining the absorber concentration from the predicted absorbance spectrum profile

Conjugated π→π* transitions, the K-bands (from the German *konjugiert*, "conjugated") typically have a molar absorption coefficient, $\varepsilon$, around $10^4$ $M^{-1}cm^{-1}$ (Kalsi, 2007). If we assume the unknown absorber with the modeled $a_{375}$ = 1278 $cm^{-1}$ (Fig. 2) is a mixture of organic molecules akin to quinine discussed above (Irvin and Irvin, 1948), there would have to be ~73 g of these organic absorbers per liter of the venusian aerosols. The modeled absorbance magnitude (without matching the $\lambda_{max}$) could be satisfied with less organic material, if we select a more efficient light absorber (summarized in Table 2). Note that, even though Table 2 presents a selection of the highly efficient UV/blue-absorbing molecules, their concentration in solution would still need to be considerable: 9.7 g/L of chlorophyll *a* (Nobel, 1999), 7.2 g/L of vitamin A (Nakahara, 2002), 4.9 g/L of beta-carotene (Nakahara, 2002), or 158 g/L of oxygenated hemoglobin protein (Prahl, 2018).

For comparison, total chlorophyll in plants typically ranges from 1 to 34 g per kg of tissue (Doddavarapu et al., 2021), while hemoglobin in human blood ranges from 120 to 175 g/L (Clark and Kruse, 1990).

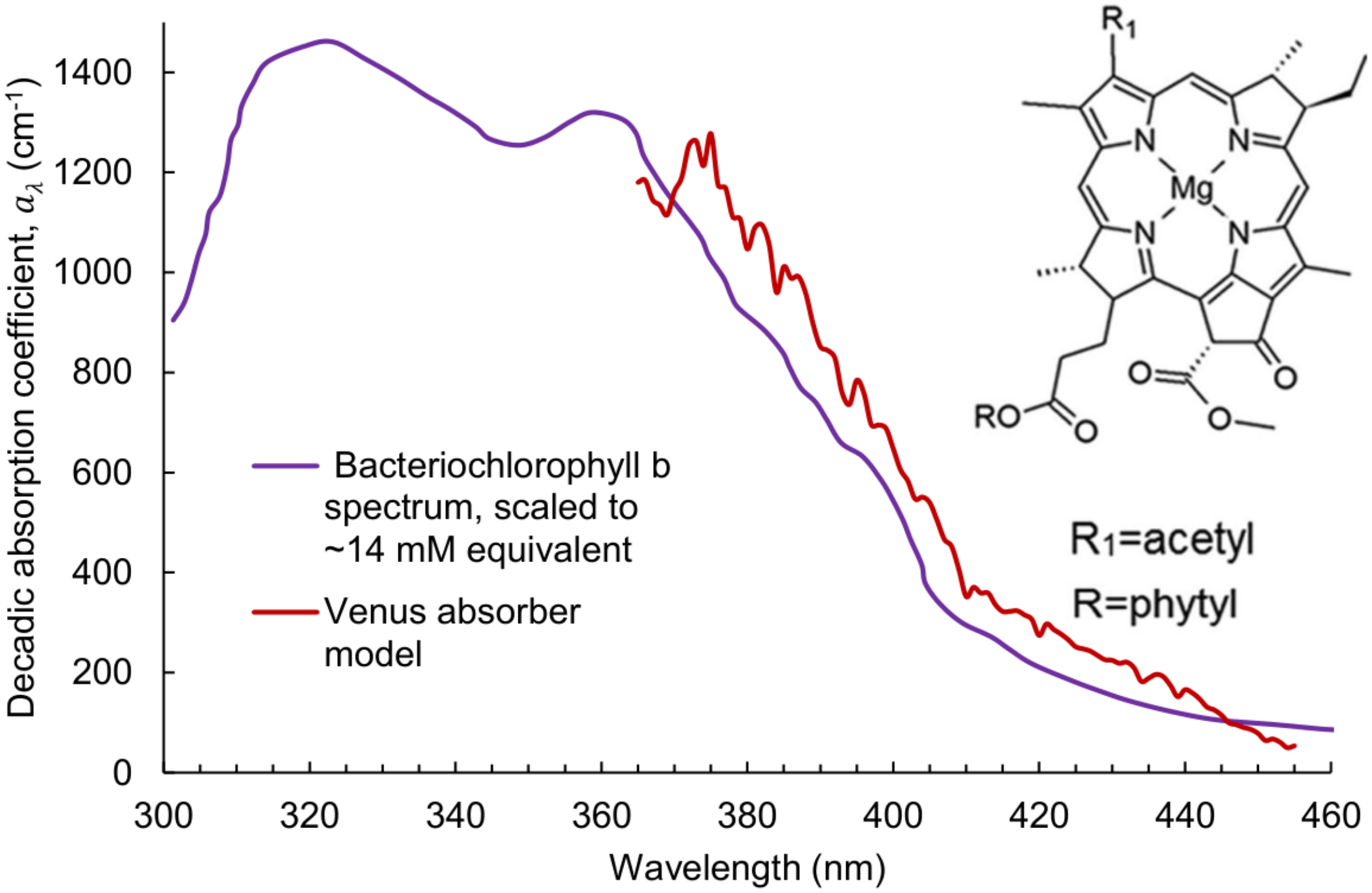

***Fig. 3.*** *Comparison of our model (red line) with an absorption spectrum profile of bacteriochlorophyll from purple bacteria Blastochloris viridis replotted from Pucelik et al. (2020) (purple line). The peak molar absorption coefficient corresponding to the published spectrum was not reported, but with a typical value 100,000 M$^{-1}$ cm$^{-1}$ for the Soret band absorption in porphyrins (Pucelik et al., 2020), the bacteriochlorophyll would have to be ~14 mM (or ~13 g/L for a molecular mass near 900 g/mol). The comparison to bacteriochlorophyll is made for illustrative purposes only, to build a reader's intuition about the required overall structural properties and the needed concentration of the light-absorbing molecules that could explain the unknown absorber in the clouds of Venus.*

We selected porphyrinoids as illustrative high-efficiency absorbers because their intense Soret bands can overlap the modeled 365–455 nm profile and they commonly have peak molar absorption coefficients on the order of 10$^5$ M$^{-1}$cm$^{-1}$. Bacteriochlorophyll b from *Blastochloris viridis* is shown in Figure 3 as one example whose spectral shape broadly resembles the modeled profile. This comparison is illustrative only, because its spectrum in concentrated sulfuric acid would likely differ from spectra measured in other solvents (see Section 2.2.3).

*Table 2. Properties of selected light-absorbing molecules and concentrations needed to produce a decadic absorption coefficient of 1278 cm$^{-1}$ at their respective $\lambda_{max}$.*

| | $\lambda_{max}$ [nm] | ε [M$^{-1}$cm$^{-1}$] | MW [g/mol] | c [g/L]* |
|---|---|---|---|---|
| Quinine | 349 | 5.7×10$^3$ | 325 | 73 |
| Vitamin A | 328 | 5.1×10$^4$ | 286.5 | 7.2 |
| Chlorophyll a | 430 | 1.2×10$^5$ | 911.5 | 9.7 |
| Beta-carotene | 451 | 1.4×10$^5$ | 536.9 | 4.9 |
| Hemoglobin $O_2$ | 414 | 5.2×10$^5$ | 64,458 | 158 |

*Needed for $a_{max}$ = 1278 cm$^{-1}$

In general, porphyrins are among the most efficient light absorbers, with molar absorption coefficients often exceeding $10^5$ $M^{-1}cm^{-1}$ at the near UV Soret band. Thus, approximately 13 mM absorber would satisfy the modeled absorption intensity. For a molecular mass near 900 g/mol, this corresponds to approximately 12 g/L of absorber. The position and overall shape of the Soret band depend on macrocycle substitution, metalation, axial ligation, molecular conformation, and solvent environment.

Note that porphyrinoids, including the example in Fig. 3, usually have a Q-band, a secondary, weaker absorption, typically between 500 and 700 nm, arising from an electronic transition with vibronic structure (Giovannetti, 2012). Since observations of Venus do not provide evidence of deep absorption longward of 455 nm that does not have a specific chemical assignment, we can exclude porphyrins with prominent Q-bands from the absorber candidate pool.

However, also note that the Q-band is absent or diminished in porphyrinoids with strong metal-to-ligand charge-transfer or in molecules with asymmetric distortions of the porphyrin macrocycle induced by axial ligands (perpendicular to the porphyrin plane) and the surrounding protein environment.

An example of a porphyrinoid chromophore without major absorption in the red part of the spectrum is heme B in oxyhemoglobin (**Table 2**) (Schweitzer-Stenner and Bigman, 2001). Oxyhemoglobin has a Soret-band maximum near 414 nm with a molar absorption coefficient of $5.2 \times 10^5$ $M^{-1}$ $cm^{-1}$ per hemoglobin tetramer, corresponding to $1.3 \times 10^5$ $M^{-1}$ $cm^{-1}$ per heme chromophore. Matching the modeled peak linear decadic absorption coefficient of 1278 $cm^{-1}$ would therefore require approximately 9.8 mmol/L heme chromophores. For comparison, the physiological concentration of hemoglobin-bound heme chromophores in human blood is approximately 8–12 mmol/L (Lodemann et al., 2010). This comparison is made at the respective Soret-band maxima and intended only to illustrate the required high chromophore molarity; hemoglobin itself does not reproduce the wavelength dependence of the Venus absorber. As noted above, Soret-band maxima vary among porphyrinoid derivatives, and a maximum near 375 nm falls within the variability range, as illustrated by the bacteriochlorophyll spectrum in Figure 3.

### 2.2.3. Other effects: Solvent, molecular interactions, chemical stability, empirical observations

The solvent has a major influence on the absorption spectra of the solutes. While absorption bands in vacuum are very sharp (with FWHM often less than a few nanometers), in solution, the absorption bands broaden due to solvent–solute interactions and vibrational coupling. Further, the peak positions can shift due to the solvatochromic effect, depending on the polarity of the solvent (Kalsi, 2007, Zuehlsdorff and Isborn, 2019).

The environment of Venus's upper cloud aerosols (approximately -30 °C and approximately 80% w/w sulfuric acid in water (Titov et al., 2018)) will likely have a major influence on the absorption band broadening and shift as compared to aqueous solutions used in the laboratory experiments. Thus, a direct comparison of an absorbance obtained from an aqueous solution of a candidate molecule to the modeled absorbance of Venus's unknown absorber has limited value if Venus's environmental conditions are not mimicked; that is, the absorber candidates' overall chemical characteristics, structure, and reactivity should be experimentally verified in cold highly concentrated sulfuric acid. The candidate absorbance in sulfuric acid will also be influenced by protonation and subsequent chemical changes caused not just by the exotic environment within the aerosol but also by various physical and chemical factors including, but not limited to, high UV radiation flux and reactions with dissolved atmospheric gases.

Another factor to consider is that molecular stacking can cause hypochromism (Kasha, 1963). Under the Beer–Lambert law, absorbance follows $A = \varepsilon c l$, increasing linearly with concentration $c$ and path length $l$ for non-interacting chromophores with constant molar absorption coefficient $\varepsilon$. Since our model favors concentrated solutions of conjugated systems (possibly planar aromatics), molecular stacking is possible; thus the concentration of organic molecules might be higher than expected from a simple linear application of the Beer–Lambert law.

In our previous work, we showed that formaldehyde dissolved in highly concentrated sulfuric acid reacts with dissolved carbon monoxide and through a cascade of reactions produces black or brown mixtures (Spacek et al., 2024) that absorb light throughout the entire UV/vis spectrum. The absorption "tail" extending from UV to infrared without significant local maxima is typical for random mixtures of conjugated molecules comparable to graphene oxide suspension (Johra et al., 2014) (Fig. 4) or crude oil (Banda-Cruz et al., 2017). This contrasts with steeply decreasing absorbance predicted by our model, which, as we discussed above, corresponds to a defined molecule or a few molecules with similar optical properties.

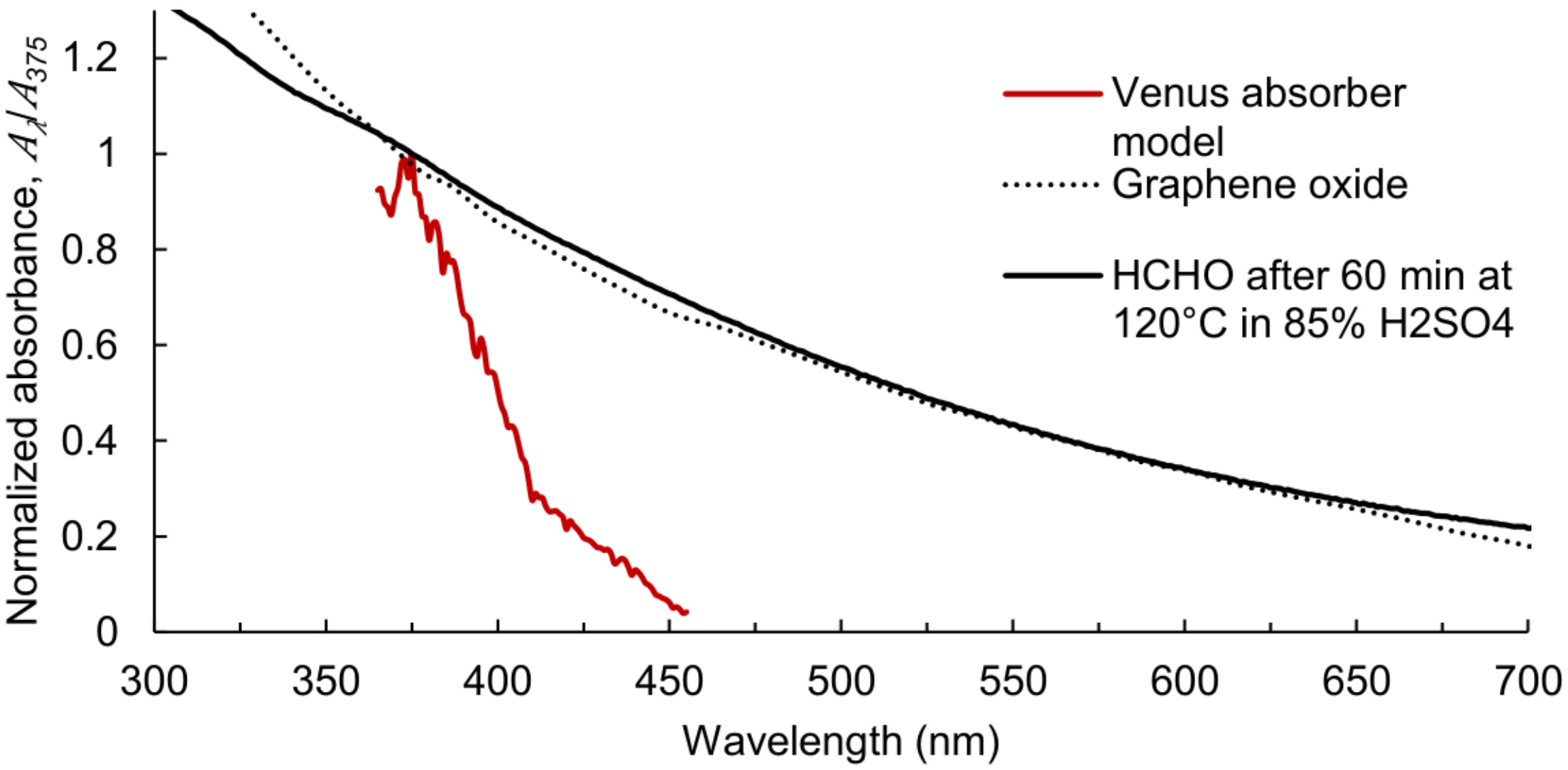

***Fig. 4.*** *Comparison of the relative absorption spectra, $A_\lambda/A_{375}$, of the Venus absorber model (red line), a mixture generated from formaldehyde (HCHO) heated for 60 min at 120 °C in 85 % w/w $H_2SO_4$ (solid black) and graphene oxide solution (dotted black). The organic tar-like mixture arising from the heated formaldehyde in sulfuric acid resembles the spectrum of crude oil or graphene oxide and on its own does not resemble the spectrum of the unknown absorber predicted by our model. Graphene oxide data were replotted from Johra et al. (2014).*

Similar tar-like products with spectra extending throughout the visible part of the spectrum were observed with most other organics we exposed to sulfuric acid (Moncada et al., 2024). As a typical example, we present glucose, which in aqueous solution is colorless (no π electrons for π→π* transitions) but after mixing with sulfuric acid undergoes immediate dehydration resulting in "rose-pink" solution (Love, 1953) with a distinct peak at 320 nm (Fig. 5). As this initial dehydration product reacts further, the absorption spectrum gradually turns to a black mixture with the typical absorbance ramp encompassing the entire absorption spectrum, suggesting a tar-like mixture production. At higher reactant concentrations (including formaldehyde) this exothermic reaction proceeds toward black solids (Spacek et al., 2024).

Figures 4 and 5 illustrate that the venusian absorber does not have a profile of a tar-like mixture. This means that if Venus's absorber is a large organic molecule, or a mixture of large organic molecules, it has to be chemically stable in highly concentrated sulfuric acid aerosols or be in a dynamic equilibrium, where the tar-like products provide negligible absorption. In other words, the spectral characteristics of the sulfuric acid solution that contains dissolved candidate organic absorber must remain unchanged.

The model presented in Spacek et al. (2024) used gas-phase abundance to estimate the steady-state concentration of dissolved formaldehyde (and its reaction products) in Venus's upper cloud aerosols to be ~0.14 mM. If the estimated 0.14 mM represents carbon equivalents, assembly into absorbers that contain approximately 50 carbon atoms would yield an absorber concentration of approximately 2.8 μM. For $\varepsilon_{375}$ = $10^5$ $M^{-1}$ $cm^{-1}$, this corresponds to $a_{375}$ ≈ 0.28 $cm^{-1}$, approximately 4560-fold below the modeled $a_{375}$ =1278 $cm^{-1}$. In other words, the model presented here indicates that if the absorber is conjugated organic carbon, it is in concentrations that far exceed predicted equilibrium with atmospheric formaldehyde and carbon monoxide (Spacek et al., 2024).

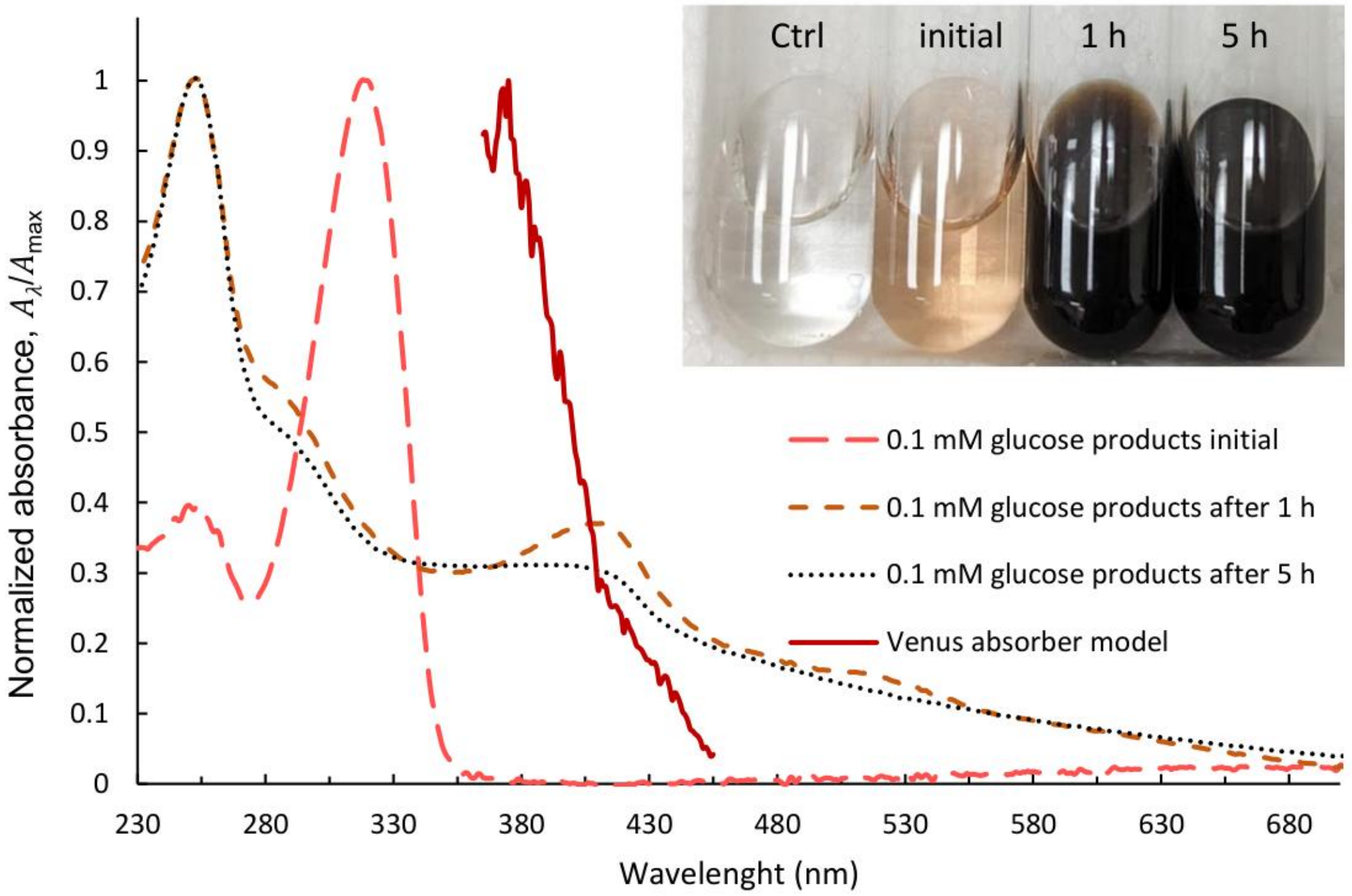

***Fig. 5.*** *Comparison of normalized spectra of glucose (itself non-absorbing) at different time points after its addition to 85% w/w sulfuric acid. The glucose reacts with sulfuric acid and within seconds produces a rose-pink solution with a single predominant product (Love, 1953), which reacts further toward the color of a tar-like mixtures. Here 0.1 mM glucose in 85% w/w sulfuric acid was heated to 60°C for the indicated time. The initial spectrum with sharply defined peaks (long dash line) was gradually converted into a broad ramp with less defined peaks (dashed and dotted lines show 1 h and 5 h heating respectively). Inset shows how the plotted progression toward tar-like mixtures appears visually using a separate, more concentrated 10 mM glucose solution. Left to right: The glucose solution in water is colorless (ctrl), but when exposed to sulfuric acid within a few seconds it turns rose-pink (initial), followed by barely transparent brown after 1 h at 60°C (see the top of the meniscus) to completely opaque solution after 5 h at 60°C.*

# 3. Discussion and Summary

As we illustrated with the "waterfall analogy" in the introduction, a cloud of small droplets appears brighter than the bulk liquid that forms it. The brightening is due to scattering at the gas–liquid interface which, for small aerosols, dominates the properties of the reflected light. Thus, a liquid that appears pitch black to the naked eye might appear "pale yellow" when it is in the form of cloud droplets, as long as the aerosols are small enough. We believe that this is the case for the venusian clouds.

It has been shown that the cloud absorber believed to reside within its cloud aerosols absorbs "about half of the solar energy the planet receives from the Sun" (Crisp, 1986, Lee et al., 2015b, Titov et al., 2018). And yet, surprisingly, we are the first to ask how "dark" the bulk liquid forming Venus's cloud is and to provide a model with an answer.

From an analytical perspective, the UV/vis absorbance spectrum of a bulk liquid constrains the properties of any absorber and allows direct quantitative comparison of proposed candidate absorbers with observations. Such analysis serves to narrow the existing pool of candidate absorbers (Limaye et al., 2021).

We note that the model is only as good as its underlying assumptions. The resulting UV/vis absorbance of the bulk liquid may vary depending on the assumed characteristics of the cloud and the unknown absorber distribution within the cloud. For example, the absorbing layer may extend upward above the modeled cloud tops if the entire upper haze layer is composed of mode 1 particles (Marcq et al., 2019, Pérez-Hoyos et al., 2018). Such a scenario would reduce the bulk-liquid absorption within the aerosols.

On the other hand, the absorber distribution might not be uniform between mode 1 and mode 2 aerosols. In this case, the particles that carry the absorber would need to be more light absorbing. The model proposed here provides the average absorption of the liquid based on 2020 Akatsuki observations (Lee et al., 2022), although there is a significant spatial and temporal variation in the observed scattered light albedo (Lee et al., 2019).

Our use of a 6-km-thick absorbing layer with the absorber well-mixed over all particle sizes provides the first conservative estimate of the model's underlying assumptions. Further study should refine these assumptions and explore the dependencies and degeneracies between observations and the resulting parameters, including (1) the vertical extent of the absorber and (2) the distribution of the absorber in the cloud particles. New observations constraining aerosol optical properties, and their spectral dependence will greatly advance further modeling efforts of the unknown absorbers within the clouds.

The future model should also investigate the spectrum across the entire range where the absorber has been documented (Pérez-Hoyos et al., 2018).

A candidate absorber should account for the observation that absorption at 283 nm is much greater than at 360 nm to reproduce the observed 283 nm phase curve. This is because greater absorbance is needed to compensate for stronger Rayleigh scattering above the absorbing layer (Lee et al., 2021). Alternatively, multiple absorbers could explain this feature if they were segregated to different altitude ranges: the strong 283 nm absorber above the cloud tops and the 360 nm absorber within the clouds themselves.

Despite the limitations of our initial model, this approach provides an empirical link between laboratory spectroscopic measurements of bulk test-tube solutions and the observed ultraviolet–visible reflectance curve of Venus. Such comparisons place limits on the intensity and the spectral features of proposed absorber candidates. Our work provides important constraints that are crucial for the identification of the unknown venusian absorber. Our model predicts that the decadic absorption coefficient of the unknown venusian absorber, as a bulk solution, is 1278 $cm^{-1}$ at 375 nm, with a moderately sharp decline to < 54 $cm^{-1}$ at 455 nm.

This is extremely high light absorption. To illustrate the magnitude of this 1278 $cm^{-1}$ absorption coefficient, we compare it with well-known examples of efficient UV/blue light absorbing molecules and show that the small-molecule examples in Table 2 require concentrations ~5–73 g/L. A chromophoric macromolecule may require a higher mass concentration, as illustrated by the hemoglobin example of 158 g/L. The ~0.5–7% w/V ratio range of small efficient organic absorbers in the aerosol is surprising yet conceivable. As an example (Table 2), we show that some porphyrin molecules might match the modeled section of the absorbance spectrum within the Soret band. We show that some porphyrins do not have a prominent band in the red part of the spectrum, for which there is currently no clear evidence in the Venus absorber. However, note that we do not propose porphyrins specifically as the venusian absorber. Indeed, it is unlikely that we could correctly guess a specific absorber from the virtually infinite possible configurations of complex organic molecules that may match the UV/vis absorption profile of the unknown venusian absorber.

Because our model omits the majority of the visible spectrum, we can give only a limited answer with regard to how the bulk liquid collected from the venusian cloud would appear to the naked eye. Based on the presented data, we would be equally unsurprised if the collected venusian cloud liquid appeared to the human eye in the range between a bright yellow paint and asphalt, depending on the absorption intensity beyond 455 nm.

However, given how little the absorbing species within aerosols contribute toward the observed cloud-top albedo compared to scattering, and due to uncertainties in the remote measurements, it is unlikely that remote observations will be able to reveal whether the decadic absorption coefficient of the bulk liquid forming Venus's clouds is below ~3 $cm^{-1}$ (transmittance 0.1% for 1 cm path length; effectively opaque to the human eye) in any part of the visible spectrum. Thus, a proper, reliable assessment of the absorbance of the bulk liquid of Venus's cloud droplets would require an *in situ* cloud probe (Baines et al., 2021, Hein et al., 2020, Seager et al., 2022b) or a dedicated atmospheric and cloud sample return mission (Agrawal et al., 2024, Schulze-Makuch et al., 2024).

This work, while constraining properties of the venusian UV–blue absorber, also raises new questions. If the modeled absorption is produced by a high-efficiency absorber ($\varepsilon \approx 1 \times 10^5$ $M^{-1}$ $cm^{-1}$ at the relevant wavelength, for example, characteristic of strong charge-transfer or $\pi \rightarrow \pi^*$ electronic transition), then it must constitute on the order of ~0.3–7% w/V of the upper-cloud aerosol (Table 2).

By contrast, an absorber with $\varepsilon_{375}$ = $10^4$ $M^{-1}$ $cm^{-1}$ would require a molar concentration of 0.128 M, while an absorber with $\varepsilon_{375}$ = $10^3$ $M^{-1}$ $cm^{-1}$ would require molar concentration of 1.28 M. Depending on molecular mass, solubility, and density, such species would have to constitute a substantial fraction of the aerosol liquid. Depending on their molecular masses, solubilities, and attainable concentrations, lower-efficiency absorber candidates may be physically or chemically implausible as explanations for the modeled absorption.

The assumed high-efficiency absorber concentration is required to be nearly four orders of magnitude greater than is the expected concentration of total organic carbon inventory in the aerosols based on the gas-phase equilibrium calculations completed by Spacek et al. (2024). The absorbance profile inferred from the observed cloud-top reflectivity is inconsistent with "tar-like mixtures" expected to be produced by an array of organic solutions when subjected to Venus's cloud conditions on long timescales (days) (Fig. 5). Such mixtures would generate absorbances in the red part of the visible spectrum far greater than those observed.

Hartley et al. (1989) proposed croconic acid as a candidate Venus cloud absorber. Colmenero and Escribano (2019) subsequently characterized the strong UV-visible absorption of croconic and related cyclic oxocarbon acids between approximately 200 and 450 nm. On this basis, these compounds merit further investigation as absorber candidates. Other abiotic pathways may exist that can generate other defined organic absorbers with the spectroscopic properties specified by the model developed here. Perhaps the physicochemical conditions in the venusian clouds (e.g. the presence of metals, concentration of acid in the aerosols, the altitude temperature gradient) can also result in a stable highly concentrated organic species formed via biotic or abiotic pathways with a defined structure that is stable and does not degrade to a tar-like substance with reflectance properties similar to that presented in Fig. 5.

Our recent results on the stability and reactivity of various classes of organic molecules show that highly concentrated sulfuric acid can support complex organic chemistry that is fundamentally structurally and functionally different from the highly cross-linked, aromatic, oxidized molecules that are identified under the umbrella term "red oil" (Duzdevich et al., 2025, Petkowski et al., 2025a, Petkowski et al., 2025b, Seager et al., 2024, Seager et al., 2023). This opens the possibility for chemically stable organic molecules to exist in the clouds of Venus and to contribute to the unknown absorber phenomenon. If the absorber is chiral organic material that produces a circular polarization in reflected light, it would be a strong indication of a biological origin (Sparks et al., 2009, Sparks et al., 2021).

If the venusian clouds are darkened by an inorganic UV/blue absorber, it must be a major component of the aerosols, as the d-d transitions in inorganic UV/blue absorbers have typically a smaller molar absorption coefficient (usually in the $10^3$ to $10^4$ $M^{-1}cm^{-1}$ range). Consider, for example, ferric chloride ($FeCl_3$). Although its absorbance spectrum corresponds well to the unknown absorber spectrum in *relative* terms (Egan et al., 2025b), its *absolute* absorbance is too low. Our model predicts $a_{375}$ = 1278 $cm^{-1}$ for the Venus cloud liquid, while an aqueous solution of 1 g/L $FeCl_3$ has $a_{375} \approx 1$ $cm^{-1}$ (Abderrazak et al., 2000). Simple linear extrapolation would therefore require approximately 1.3 kg/L $FeCl_3$ to match the modeled coefficient.

The possibility that Venus's cloud is darkened through $\pi \rightarrow \pi^*$ transitions in organic absorber(s) motivated us to design space missions that directly probe the cloud particles for the presence of organic material. Our laboratory results indicate that a 440 nm excitation laser would be able to excite fluorescence in most organic solutions generated in sulfuric acid (Moncada et al., 2024). These solutions would produce fluorescence emission detectable around 500 nm, which lies in the more transparent region of the modeled venusian absorber and matches the detector window of the autofluorescence nephelometer (AFN) instrument (Baumgardner et al., 2022). No known inorganic species present in the clouds of Venus could provide such a fluorescence signal. Hence the detection of 500 nm fluorescence will be a strong indication of the presence of large, conjugated organics in Venus's cloud aerosols.

The planned Rocket Lab mission to Venus (French et al., 2022) is designed to deliver a probe containing the AFN instrument to search for fluorescence and to observe backscattered polarized radiation in Venus's cloud droplets (Baumgardner et al., 2022). Detection of the fluorescence signal would be indicative of organic molecules.

More sophisticated future missions, planned as a part of the "Morning Star Missions to Venus" initiative (Seager et al., 2022a, Seager et al., 2022b), go beyond the mere detection of organics in the cloud droplets and aim to identify organic species, if they are present (Ligterink et al., 2022).

# Acknowledgements

We thank Breakthrough Initiatives for partial funding of this work. We acknowledge useful discussions with the Morning Star Missions to Venus team, led by Sara Seager, and thank Steven Benner of the Foundation for Applied Molecular Evolution for reviewing the manuscript.

# Author Contributions

J.S. conceived the original idea, wrote the manuscript, and prepared figures. Y.J.L. developed the model, prepared s, and contributed to manuscript editing. P.B.R. reviewed the model and contributed to manuscript editing. J.J.P. contributed to manuscript writing and editing.

# Conflict of interest

The authors declare no conflict of interest.

# Funding

This research was partially funded by Breakthrough Initiatives. Y.J.L. was supported by the Institute for Basic Science (IBS-R035-C1).